\documentclass[twocolumn,aps,prl,showpacs,superscriptaddress]{revtex4}
\usepackage{graphicx}
\usepackage{amsfonts}
\usepackage{amsmath}
\usepackage{amssymb}

\begin{document}


\title{Experimental Fully Contextual Correlations}




\author{Elias Amselem}
 \affiliation{Department of Physics, Stockholm University, S-10691
 Stockholm, Sweden}
\author{Lars Eirik Danielsen}
 \affiliation{Department of Informatics,
 University of Bergen, P.O. Box 7803, Bergen N-5020, Norway}
\author{Antonio J. L\'{o}pez-Tarrida}
 \affiliation{Departamento de F\'{\i}sica Aplicada II,
 Universidad de Sevilla, E-41012 Sevilla, Spain}
\author{Jos\'{e} R. Portillo}
 \affiliation{Departamento de Matem\'{a}tica Aplicada I,
 Universidad de Sevilla, E-41012 Sevilla, Spain}
\author{Mohamed Bourennane}
 \affiliation{Department of Physics, Stockholm University, S-10691
 Stockholm, Sweden}
\author{Ad\'an Cabello}
 \affiliation{Departamento de F\'{\i}sica Aplicada II, Universidad de
 Sevilla, E-41012 Sevilla, Spain}
 \affiliation{Department of Physics, Stockholm University, S-10691
 Stockholm, Sweden}


\date{\today}



\begin{abstract}
Quantum correlations are contextual yet, in general, nothing prevents the existence of even more contextual correlations. We identify and test a
noncontextuality inequality in which the quantum violation cannot be improved by any hypothetical postquantum theory, and use it to experimentally obtain correlations in which the fraction of noncontextual correlations is less than $0.06$. Our correlations are experimentally generated from the results of sequential compatible tests on a four-state quantum system encoded in the polarization and path of a single photon.
\end{abstract}




\maketitle


{\em Introduction.---}Quantum contextuality \cite{Specker60,Bell66,KS67} refers to the fact that the predictions of quantum mechanics (QM) cannot be reproduced assuming noncontextuality of results (i.e., that the results are predefined and independent of other compatible tests) or, equivalently, noncontextual hidden variable theories. By compatible tests we mean those satisfying the following theory-independent definition: ``If a physical system is prepared in such a way that the result of test $x_i$ is predictable and repeatable, and if a {\em compatible} test $x_j$ is then performed (instead of test $x_i$) a subsequent execution of test $x_i$ shall yield the same result as if test $x_j$ had not been performed'' \cite{Peres95} (see \cite{GKCLKZGR10} for other definitions of compatibility). In QM, two tests represented by self-adjoint operators $A$ and $B$ are compatible when $A$ and $B$ commute. This guarantees that the quantum predictions for compatible tests are given by a single probability measure on a single probability space. Compatibility implies that the probability $P(a_i|x_i)$ of obtaining the result $a_i$ for the test $x_i$ is independent of other compatible tests $x_1,\ldots,x_{i-1},x_{i+1},\ldots,x_n$, i.e.,
\begin{equation}
 P(a_i|x_i)=\sum_{a_1,\ldots,a_{i-1},a_{i+1},\ldots,a_n} P(a_1,\ldots,a_n | x_1,\ldots, x_n),
 \label{compatibility}
\end{equation}
for all sets $x_1,\ldots,x_n$ of compatible tests, and where $P(a_1,\ldots, a_n|x_1,\ldots, x_n)$ is the joint probability of obtaining the results $a_1,\ldots,a_n$ for the compatible tests $x_1,\ldots,x_n$, respectively. Assumption \eqref{compatibility} is formally equivalent to the no-signaling principle, but involves compatible tests instead of spacelike separated tests.

The assumption of the noncontextuality of results states that the result $a_i$ of test $x_i$ is the same regardless of other compatible tests being performed; it only depends on $x_i$ and some hidden variables $\lambda$. This implies that the correlation among the results of compatible tests can be expressed as
\begin{equation}
 P(a_1,\ldots, a_n|x_1,\ldots, x_n) = \sum_{\lambda} P(\lambda) \prod_{i=1}^n P(a_i|x_i,\lambda),
 \label{noncontextuality}
\end{equation}
for some common distribution $P(\lambda)$.

Noncontextuality inequalities are expressions of the form
\begin{equation}
 S \equiv \sum T_{a_1,\ldots, a_n, x_1,\ldots, x_n}P(a_1,\ldots, a_n|x_1,\ldots, x_n) \stackrel{\mbox{\tiny{ NC}}}{\leq} \Omega_\mathrm{NC},
 \label{nci}
\end{equation}
where $T_{a_1,\ldots, a_n, x_1,\ldots, x_n}$ are real numbers and $\stackrel{\mbox{\tiny{ NC}}}{\leq} \Omega_\mathrm{NC}$ denotes that the maximum value of $S$ for any noncontextual correlations [therefore satisfying \eqref{noncontextuality}] is $\Omega_\mathrm{NC}$. Quantum contextuality is experimentally observed through the violation of noncontextuality inequalities \cite{KZGKGCBR09,MWZ00,BKSSCRH09,ARBC09}.

Quantum nonlocality \cite{Bell64} is a particular form of quantum contextuality which occurs when the tests are not only compatible but also spacelike separated. In this case, noncontextuality inequalities are called Bell inequalities \cite{Bell64}. In addition to applications such as device-independent quantum key distribution \cite{BHK05,ABGMPS07} and random number generation \cite{PAMBMMOHLMM10}, which require spacelike separation, quantum contextuality also offers advantages in scenarios without spacelike separation. Examples are communication complexity \cite{BCMD10}, parity-oblivious multiplexing \cite{SBKTP09}, zero-error classical communication \cite{CLMW10}, and quantum cryptography secure against specific attacks \cite{Svozil10,CDNS11}.

The goal of this work is to identify and perform an experiment with sequential quantum compatible tests, which produces correlations with the largest contextuality allowed under the assumption \eqref{compatibility}, which is assumed to be valid also for postquantum theories. For this purpose, we first introduce a measure of contextuality of the correlations, the noncontextual content $W_\mathrm{NC}$, so that $W_\mathrm{NC}=0$ corresponds to the maximum contextuality. Then, we show how to experimentally obtain testable upper bounds to $W_\mathrm{NC}$. Next, we show how graph theory allows us to identify experiments in which the upper bound to $W_\mathrm{NC}$ predicted by QM is zero, and apply this method to single out an experiment for which $W_\mathrm{NC}=0$. Finally, we perform this experiment and obtain correlations in which $W_\mathrm{NC}<0.06$.


{\em Noncontextual content.---}Every correlation among compatible tests [therefore satisfying \eqref{compatibility}] can be expressed as
\begin{equation}
\label{decomposition}
 \begin{split}
 P(a_1,\ldots, a_n|x_1,\ldots, x_n)=
 w_\mathrm{NC} P_\mathrm{NC}(a_1,\ldots, a_n|x_1,\ldots, x_n) \\
 +
 (1-w_\mathrm{NC}) P_\mathrm{C}(a_1,\ldots, a_n|x_1,\ldots, x_n),
 \end{split}
\end{equation}
where $0 \leq w_\mathrm{NC} \leq 1$, $P_\mathrm{NC}(a_1,\ldots, a_n|x_1,\ldots, x_n)$ can be expressed as \eqref{noncontextuality}, and $P_\mathrm{C}(a_1,\ldots, a_n|x_1,\ldots, x_n)$ satisfies \eqref{compatibility} but cannot be expressed as \eqref{noncontextuality}. We define the {\em noncontextual content} $W_\mathrm{NC}$ of the correlations as the maximum value of $w_\mathrm{NC}$ over all possible decompositions as \eqref{decomposition}, i.e.,
\begin{equation}
 W_\mathrm{NC} \equiv \max_{\{P_\mathrm{NC},P_\mathrm{C}\}}w_\mathrm{NC}.
\end{equation}
This definition is parallel to the definition of local content introduced in \cite{BKP06}. In fact, for correlations generated through spacelike separated tests, the noncontextual content equals the local content.

$\Omega_\mathrm{NC}$, $\Omega_\mathrm{Q}$, and $\Omega_\mathrm{C}$ will denote, respectively, the maximum value of $S$ for noncontextual correlations [i.e., which can be expressed as \eqref{noncontextuality}], quantum correlations, and correlations satisfying \eqref{compatibility}. Now consider correlations satisfying \eqref{compatibility} and saturating $\Omega_\mathrm{Q}$. Then, given a decomposition of such correlations as \eqref{decomposition}, with $w_\mathrm{NC}=W_\mathrm{NC}$, $\Omega_\mathrm{Q}$ can be expressed as $\Omega_\mathrm{Q}=\sum T_{a_1,\ldots, a_n, x_1,\ldots, x_n}[W_\mathrm{NC} P_\mathrm{NC}(a_1,\ldots, a_n|x_1,\ldots, x_n)+(1-W_\mathrm{NC}) P_\mathrm{C}(a_1,\ldots, a_n|x_1,\ldots, x_n)]=W_\mathrm{NC} \sum T_{a_1,\ldots, a_n, x_1,\ldots, x_n} P_\mathrm{NC}(a_1,\ldots, a_n|x_1,\ldots, x_n)+(1-W_\mathrm{NC}) \sum T_{a_1,\ldots, a_n, x_1,\ldots, x_n} P_\mathrm{C}(a_1,\ldots, a_n|x_1,\ldots, x_n)$. The first sum can be expressed in a noncontextual form, so it is upper bounded by $\Omega_\mathrm{NC}$. The second sum cannot be expressed in a noncontextual form, so it can only be upper bounded by $\Omega_\mathrm{C}$. Hence, $\Omega_\mathrm{Q} \le W_\mathrm{NC} \Omega_\mathrm{NC} + (1-W_\mathrm{NC}) \Omega_\mathrm{C}$, and, taking into account that $\Omega_\mathrm{NC} \le \Omega_\mathrm{Q} \le \Omega_\mathrm{C}$, then
\begin{equation}
 \label{bound}
 W_\mathrm{NC}\leq\frac{\Omega_\mathrm{C}-\Omega_\mathrm{Q}}{\Omega_\mathrm{C}-\Omega_\mathrm{NC}}.
\end{equation}
Any experimental violation $S_\mathrm{exp}$ of a noncontextuality inequality indicates that $\Omega_\mathrm{C} > \Omega_\mathrm{NC}$ and, therefore, provides an upper bound on $W_\mathrm{NC}$, namely $W_\mathrm{NC} \leq (\Omega_\mathrm{C}-S_\mathrm{exp})/(\Omega_\mathrm{C}-\Omega_\mathrm{NC})$. Assuming that the maximum $S_\mathrm{exp}$ in an ideal experiment is given by $\Omega_\mathrm{Q}$, to observe correlations with zero noncontextual content, here called {\em fully contextual correlations}, one has to test a noncontextuality inequality such that its maximum quantum violation equals its maximum possible violation under the assumption \eqref{compatibility}, i.e., an inequality for which $\Omega_\mathrm{NC}<\Omega_\mathrm{Q}=\Omega_\mathrm{C}$.

However, even if $\Omega_\mathrm{Q}=\Omega_\mathrm{C}$, inherent imperfections of actual experiments will prevent the observation of $W_\mathrm{NC}=0$. In general, the more complex the experiment to produce the required quantum correlations is, the higher the probability that experimental imperfections lead to a higher upper bound for the noncontextual content. Therefore, the task is to identify the simplest noncontextuality inequality violated by QM and such that $\Omega_\mathrm{Q}=\Omega_\mathrm{C}$.


\begin{figure}[t]
\centerline{\includegraphics[width=8cm]{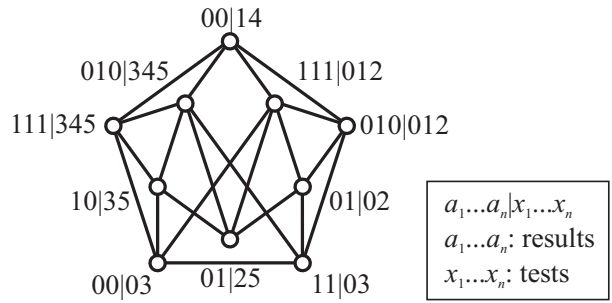}}
\caption{\label{Fig1} Graph corresponding to inequality (\ref{inequality}). Vertices represent propositions. For example, $01|25$ means ``result $0$ is obtained when test $2$ is performed, and result $1$ is obtained when test $5$ is performed.'' Edges link propositions that cannot be simultaneously true. For example, $01|25$ and $01|02$ are linked, since in the first proposition the result of test $2$ is $0$, while in the second proposition the result of test $2$ is $1$.}
\end{figure}


{\em Graph approach.---}We addressed this problem by using a connection between graph theory and noncontextuality inequalities noticed in \cite{CSW10}: For any graph there is a noncontextuality inequality for which $\Omega_\mathrm{NC}$, $\Omega_\mathrm{Q}$, and $\Omega_\mathrm{C}$ are given, respectively, by the independence number, the Lov\'asz number, and the fractional packing number of the graph \cite{SM}. We calculated these three numbers for all nonisomorphic graphs with less than 11 vertices, and found that there are no graphs with less than 10 vertices with $\Omega_\mathrm{NC}<\Omega_\mathrm{Q}=\Omega_\mathrm{C}$, and there are only four 10-vertex graphs with these properties \cite{SM}. The maximum quantum violation of noncontextuality inequalities associated with three of them requires quantum systems of dimension higher than four, while dimension four is enough for the graph in Fig.~\ref{Fig1}. The inequality associated with the graph is constructed by looking for propositions involving compatible tests, such that each vertex represents one proposition in the inequality and the edges only link propositions that cannot be simultaneously true. Then, the inequality is simply given by the sum of all the probabilities of the propositions represented in the graph.

For the graph in Fig.~\ref{Fig1}, it can be easily seen that the following noncontextuality inequality is in one-to-one correspondence with the graph:
\begin{equation}
\label{inequality}
\begin{split}
S\equiv &P(010|012)+P(111|012)+P(01|02)+P(00|03)\\
 &+P(11|03)+P(00|14)+P(01|25)+P(010|345)\\
 &+P(111|345)+P(10|35) \stackrel{\mbox{\tiny{NC}}}{\leq} 3,
\end{split}
\end{equation}
where $P(10|35)$ is the probability of obtaining result $1$ when test $3$ is performed and result $0$ when test $5$ is performed. In this case, the coefficients $T_{a_1,\ldots, a_n, x_1,\ldots, x_n}$ in \eqref{nci} are all $1$. The noncontextual bound, $\Omega_\mathrm{NC}=3$, can be obtained from the independence number of the graph in Fig.~\ref{Fig1}. The maximum quantum violation of inequality \eqref{inequality} and its maximum possible violation under the assumption \eqref{compatibility} can be obtained from the Lov\'asz and the fractional packing numbers of the graph in Fig.~\ref{Fig1}, respectively \cite{SM}. This gives
\begin{equation}
 \Omega_\mathrm{Q}=\Omega_\mathrm{C}=3.5.
 \label{predictions}
\end{equation}
The maximum quantum violation can be achieved by preparing a four-state quantum system in the state
\begin{equation}
 |\psi\rangle = \frac{1}{\sqrt{2}} \left(|0\rangle + |3\rangle\right),
 \label{state}
\end{equation}
where $\langle 0 |=(1,0,0,0)$, $\langle 1 |=(0,1,0,0)$, $\langle 2 |=(0,0,1,0)$, and $\langle 3 |=(0,0,0,1)$, and with the tests represented by the following tensor products of Pauli matrices $\sigma_i$ and the $2 \times 2$ identity matrix $\openone$:
\begin{eqnarray}
 0=\sigma_x \otimes \openone,\;\;\;\;&1=\openone \otimes \sigma_z,\;\;\;\;&2=\sigma_x \otimes \sigma_z, \nonumber\\
 3=\openone \otimes \sigma_x,\;\;\;\;&4=\sigma_z \otimes \openone,\;\;\;\;&5=\sigma_z \otimes \sigma_x.
 \label{measurements}
\end{eqnarray}
The results $0$ and $1$ correspond to the eigenvalues $-1$ and $+1$, respectively, of the operators in \eqref{measurements}. Notice that every probability in \eqref{inequality} includes only pairs or trios of mutually compatible tests.

\begin{figure}[t]
\centerline{\includegraphics[width=8.4cm]{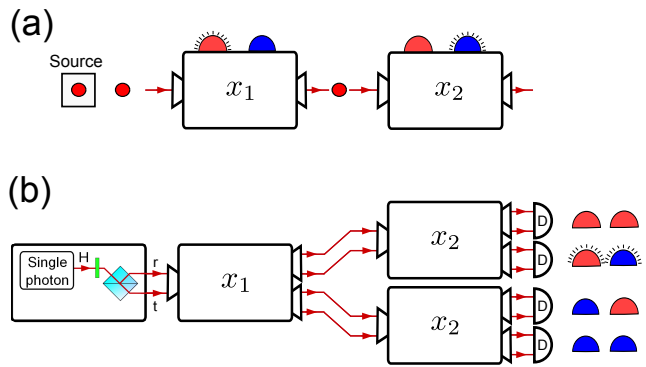}}
\caption{\label{Fig2} (color online) (a) Scheme for sequential tests of $x_{1}$ and $x_{2}$. The two possible results of each test are assigned the values $+1$ and $-1$, and are represented by whichever lamp is flashing. (b) Cascade setup used to implement two sequential tests on a single photon. It consists on three parts: state preparation, testing devices, and detectors. The preparation part produces the polarization-spatial path-encoded single-photon state $|\psi \rangle$. The two outputs of the device for testing $x_{1}$ correspond to the two possible results. After each of these two outputs we placed a device for testing $x_{2}$. Single photon detectors are placed at each of the four outputs of the two devices $x_{2}$ (see the main text for details).}
\end{figure}

{\em Experiment.---}The experiment required two-test sequences [for instance, to obtain $P(00|14)$], and three-test sequences [for instance, to obtain $P(010|012)$]. We built six devices for the six dichotomic tests defined in \eqref{measurements}. The sequential tests were performed using cascade setups \cite{ARBC09} like the one shown in Fig.~\ref{Fig2}. We tested inequality \eqref{inequality} using the
spatial path and polarization of a single photon carrying a four-state quantum system with the following encoding:
\begin{equation}
 |0\rangle=|t, H\rangle,\;\;|1\rangle=|t, V\rangle,\;\;|2\rangle=|r, H\rangle,\;\;|3\rangle=|r, V\rangle,
\end{equation}
where $t$, $r$, $H$, and $V$ denote the transmitted path, reflected path, horizontal, and vertical polarization of the photon, respectively.

The cascade setup used to implement two sequential tests on a single photon consists of three parts: state preparation, testing devices, and detectors. The preparation of the polarization-spatial path-encoded single photon state $|\psi \rangle$ is achieved using a source of $H$-polarized single photons. This single-photon source consists on an attenuated stabilized narrow bandwidth diode laser emitting at the wavelength of 780 nm. This laser offers a long coherence length. The two-photon coincidences were set to a negligible level by attenuating the laser to a mean photon number of $0.06$ per time coincidence window. This source is followed by a half-wave plate (HWP) set at $22.5^{\circ}$ and a polarizing beam splitter (PBS), allowing the photon to be distributed with equal probability between the two paths $t$ and $r$ with the right polarization $H$ and $V$, respectively [see Fig.~\ref{Fig2}].


Then, the photon in the two paths enters the device for testing $x_{1}$ through the device's input and follows one of the two possible outputs, which correspond to the values $+1$ and $-1$. After each of the two outputs we placed a device for testing $x_{2}$. We used two identical devices for testing $x_{2}$. Finally, we placed a single-photon detector (D) at the output of the two devices $x_{2}$. The same idea is used for sequences of three tests $x_{1}$, $x_{2}$, and $x_{3}$, by adding four devices for measuring $x_{3}$ and using eight single-photon detectors.

\begin{figure}[t]
\centerline{\includegraphics[width=8.4cm]{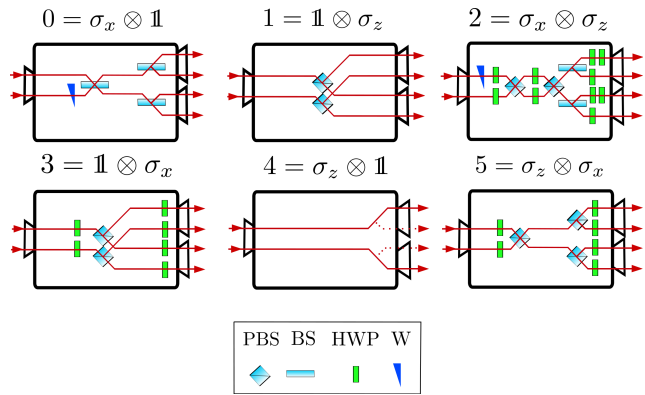}}
\caption{\label{Fig3} (color online) Devices for measuring the six tests defined in \eqref{measurements}. The technique used consists on mapping the eigenstates of the operator to the two states $|t, \phi \rangle$ and $|r, \phi \rangle$, where $\phi$ is a polarization state (see the main text for details).}
\end{figure}

Devices for measuring the six tests defined in \eqref{measurements} are given in Fig.~\ref{Fig3}. Measurements $1$ and $3$ are standard polarization measurements using a PBS and a HWP which map the polarization eigenstate of the operator to $|t, H \rangle$ and $|r,V \rangle$. The mapping to the eigenstates of test $0$, namely $(|t\rangle \pm |r\rangle)/ \sqrt{2}$, was accomplished by interfering the two paths in a $50/50$ beam splitter (BS). A wedge (W) is placed in one of the paths to set the phase between both paths [see Fig.~\ref{Fig3}]. Tests $2$ and $5$ are represented by the tensor product of a spatial path and a polarization operator so they have a four-dimensional eigenspace. However, since the tests need to be rowwise and columnwise compatible, only their
common eigenstates can be used for distinguishing the eigenvalues. Measurement $4$ requires us only to distinguish between paths $t$ and $r$. We needed to recreate the eigenstates of the performed tests after each mapping and before entering the next test, since our single-test devices map eigenstates to a fixed spatial path and polarization.


All interferometers in the experimental setup were based on a displaced Sagnac configuration. The stability of these interferometers is very high. We obtained visibilities over $99\%$ for phase insensitive interferometers, and ranging between $90\%$ and $95\%$ for phase sensitive interferometers. We used silicon avalanche photodiodes calibrated to have the same detection efficiency for single-photon detection. All single counts were registered using an eight-channel coincidence logic with a time window of $1.7$ ns. The raw detection events were gathered in a 10-second time period for each of the six experimental configurations.

The experimental results are presented in Table \ref{TableI}. The errors in the results were deduced from the standard deviation of 50 samples in the 10-second time period. The main sources of systematic errors were the small imperfections in the interferometers and in the overlapping of the light modes and the polarization components. These are the causes of the deviation of the experimental results from the ideal case observed in Table \ref{TableI}. The fact that some of the experimental results exceeded the corresponding ideal predictions was due to the lack of perfect compatibility between the sequential tests caused by the nonperfect visibilities of the interferometers. Reference \cite{GKCLKZGR10} explains how to deal with this loophole.

From the results in Table \ref{TableI}, we can establish the following experimental upper bound to the noncontextual content of the correlations:
\begin{equation}
 W_\mathrm{NC} \leq 0.0658 \pm 0.0019.
\end{equation}
This is the lowest experimental bound on the noncontextual content ever reported in any Bell or noncontextuality inequalities experiment. The previous lowest experimental upper bound on the noncontextual (local) content was $0.218 \pm 0.014$ \cite{AGACVMC11}.

As in most experiments of Bell and noncontextuality inequalities with photons, we assumed that the detected photons were an unbiased sample of the prepared photons. This assumption is necessary, since the detection efficiency, without taking into account the losses in the setup, was $0.50$ (a value obtained considering that the detection efficiency of the single-photon detectors was $55\%$ and the efficiency of the fiber coupling was $90\%$). Future experiments using heralded sources and single-photon detectors of very high efficiency \cite{LCPMN10,FFNAYTFIIIZ11} may close this loophole. Our experiment was intended to be a proof-of-principle experiment to illustrate the power of the graph approach \cite{CSW10} to single out experiments with properties on demand (in our case, $\Omega_\mathrm{NC}<\Omega_\mathrm{Q}=\Omega_\mathrm{C}$), and to experimentally observe fully contextual correlations.


\begin{table}[t]
\caption{\label{TableI}Experimental results for inequality \eqref{inequality}. The column ``Ideal'' refers to the predictions
of QM for an ideal experiment.}
\begin{center}
\begin{ruledtabular}{
\begin{tabular}{l c l}
Probability & Experimental result & Ideal\\
\hline
 $P(010|012)$ & $0.24091 \pm 0.00021$ & $0.25$ \\
 $P(111|012)$ & $0.30187 \pm 0.00020$ & $0.25$ \\
 $P(01|02)$ & $0.28057 \pm 0.00020$ & $0.25$ \\
 $P(00|03)$ & $0.50375 \pm 0.00014$ & $0.5$ \\
 $P(11|03)$ & $0.47976 \pm 0.00014$ & $0.5$ \\
 $P(00|14)$ & $0.47511 \pm 0.00034$ & $0.5$ \\
 $P(01|25)$ & $0.43765 \pm 0.00015$ & $0.5$ \\
 $P(010|345)$ & $0.24296 \pm 0.00051$ & $0.25$ \\
 $P(111|345)$ & $0.25704 \pm 0.00052$ & $0.25$ \\
 $P(10|35)$ & $0.24751 \pm 0.00035$ & $0.25$ \\
 $\Omega$ & $3.4671 \pm 0.0010$ & $3.5$ \\
\end{tabular}
} \end{ruledtabular}
\end{center}
\end{table}


{\em Conclusions.---}By using a new technique based on graph theory \cite{CSW10}, we have identified and performed an experiment in which no hypothetical postquantum correlations satisfying \eqref{compatibility} can outperform the contextuality of quantum correlations. Assuming that the detected photons are a fair sample of those emitted by the source and assuming that the compatibility of the sequential tests is perfect, the correlations observed in our experiment exhibit the largest contextuality ever reported in any experiment of Bell or noncontextuality inequalities, and provide compelling evidence of the existence of fully contextual correlations (i.e., those without noncontextual content) in nature.

Moreover, we have demonstrated the usefulness of the approach to quantum correlations based on graph theory \cite{CSW10} in identifying experiments with properties on demand. We expect that further developments along these lines will provide better tools to identify and observe phenomena of physical interest.


\begin{acknowledgments}
The authors thank M. R{\aa }dmark for his help during the experiment, and A. Ac\'{\i}n, L. Aolita, C. Budroni, R. Gallego, P. Mataloni, S. Severini, G. Vallone, and A. Winter, for stimulating discussions. This work was supported by the Swedish Research Council (VR), the Research Council of Norway, the Spanish Projects Nos.\ FIS2008-05596, MTM2008-05866, and FIS2011-29400, and the Wenner-Gren Foundation.
\end{acknowledgments}


\section{Supplemental Material}


{\em Definitions.---}In \cite{CSW10} it is shown that any connected graph $G$ can be associated to a noncontextual inequality such that: (i) its noncontextual bound $\Omega_\mathrm{NC}$ is given by the independence number $\alpha(G)$, (ii) its maximum quantum value $\Omega_\mathrm{Q}$ is given by the Lov\'asz number $\vartheta(G)$, and (iii) its maximum value for general theories satisfying that the sum of probabilities of mutually exclusive propositions cannot be larger than 1, $\Omega_\mathrm{C}$, is given by the fractional packing number $\alpha^*(G)$. The definitions follow:

The independence number $\alpha(G)$ is the maximum number of pairwise nonlinked vertices \cite{Diestel10}.

The Lov\'asz number \cite{Lovasz79} is
\begin{equation}
\vartheta(G) = \max \sum_{i=1}^{n}
|\langle\psi|v_{i}\rangle|^{2},
\end{equation}
where the maximum is taken over all unit vectors $|\psi\rangle$ and $|v_{i}\rangle$, where each $|v_{i}\rangle$ corresponds to a vertex of $G$ and two vertices are linked if and only if the vectors are orthogonal. The set $\{|v_{i}\rangle\}$ provides an orthogonal representation of the complement of $G$ (the graph such that two vertices are adjacent if and only if they are not adjacent in $G$).

The fractional packing number \cite{SU97} is
\begin{equation}
\alpha^* (G) = \max
\sum_{i\in V} w_i,
\end{equation}
where $V$ is the set of vertices of $G$, and the maximum is taken for all $0 \leq w_i\leq 1$ and for all cliques $c_j$ (subsets of mutually linked vertices) of $G$, under the restriction $\sum_{i \in c_j} w_i \leq 1$.


{\em Methods.---}We generated all nonisomorphic graphs with less than 11 vertices using {\tt nauty} \cite{McKay90}. There are 11989764 of them. For each of them we calculated $\alpha(G)$ using {\tt Mathematica} \cite{Mathematica} and $\vartheta(G)$ using {\tt SeDuMi} \cite{SeDuMi} and {\tt DSDP} \cite{Benson, BYZ00}. There are 992398 graphs for which $\alpha(G) < \vartheta(G)$. Then, we calculated $\alpha^*(G)$ using {\tt Mathematica} from the clique-vertex incidence matrix of $G$ obtained from the adjacency matrix of $G$ using {\tt MACE} \cite{MACE, MU04}, an algorithm for enumerating all maximal cliques. There are only four graphs for which $\alpha(G) < \vartheta(G)=\alpha^*(G)$; all of them have 10 vertices. The minimum dimension of the quantum system needed for the maximum quantum violation is given by the minimum dimension of the orthogonal representation of the complement of the graph leading to $\vartheta(G)$. Using this, it can be shown that only the complement of the graph in Fig.\ 1 admits an orthogonal representation in dimension four. A list containing $G$, $\alpha(G)$, $\vartheta(G)$, and $\alpha^*(G)$ for all graphs with less than 11 vertices for which $\alpha(G) < \vartheta(G)$ is provided in \cite{Web}.



\end{document}